\documentclass[twocolumn,preprintnumbers,amsmath,amssymb
prb,
]{revtex4}

\usepackage{graphicx}
\usepackage{dcolumn}
\usepackage{bm}
\usepackage{euscript}
\usepackage{float}
\usepackage[colorlinks=true]{hyperref}
\usepackage{cancel}

\newcommand{\nn}{\nonumber}
\newcommand{\ts}{\textstyle}

\renewcommand{\ln}{\mathrm{ln}}

\begin{document}

\title{Quantum fluctuations and phase coherence in superconducting nanowires}

\author{Alexey Radkevich}
\affiliation{I.E. Tamm Department of Theoretical Physics, P.N. Lebedev Physical Institute, 119991 Moscow, Russia}

\author{Andrew G. Semenov}
\affiliation{I.E. Tamm Department of Theoretical Physics, P.N. Lebedev Physical Institute, 119991 Moscow, Russia}
\affiliation{National Research University Higher School of Economics, 101000 Moscow, Russia}

\author{Andrei D. Zaikin}
\affiliation{Institute of Nanotechnology, Karlsruhe Institute of Nanotechnology (KIT), 76021, Karlsruhe, Germany}
\affiliation{National Research University Higher School of Economics, 101000 Moscow, Russia}

\date{\today}

\begin{abstract}

Quantum behavior of superconducting nanowires may essentially depend on the employed experimental setup. Here we investigate a setup that enables passing equilibrium supercurrent across an arbitrary segment of the wire without restricting fluctuations of its superconducting phase. The low temperature physics of the system is determined by a combined effect of collective sound-like plasma excitations and quantum phase slips. At $T=0$ the wire exhibits two quantum phase transitions, both being controlled by the dimensionless wire impedance $g$. While thicker wires with $g>16$ stay superconducting, in thinnest wires with $g<2$ the supercurrent is totally destroyed by quantum fluctuations. The intermediate phase with $2<g<16$ is characterized by two different correlation lengths demonstrating superconducting-like behavior at shorter scales combined with vanishing superconducting response in the long scale limit.

\end{abstract}
\maketitle

\section{\label{intro} Introduction}

Superconducting properties of quasi-one-dimensional metallic structures may dramatically differ from those of bulk systems due to pronounced fluctuation effects \cite{AGZ,book} which may persist down to lowest temperatures $T \to 0$. An important parameter that controls the magnitude of such effects in ultrathin superconducting wires is dimensionless conductance $g_\xi =R_q/R_\xi$, where $R_q =2\pi/e^2 \simeq 25.8$ K$\Omega$ is the quantum resistance unit and $R_\xi$ is the normal state resistance of the wire segment of length equal to the superconducting coherence length $\xi$. This parameter $g_\xi$ is related to the so-called Ginzburg number $Gi$ in one dimension \cite{LV} as  $g_\xi \sim Gi_{1D}^{-3/2}$ and it accounts both for small (Gaussian) and large (non-Gaussian) fluctuations of the superconducting order parameter $\Delta$. 

Gaussian fluctuations are known to yield a negative correction to the mean field value of the order parameter $\Delta\to \Delta -\delta \Delta$ with \cite{GZ08} $\delta \Delta  \sim \Delta /g_\xi $. Non-Gaussian fluctuations -- the so-called phase slips  -- correspond to temporal local suppression of the superconducting order parameter inside the wire accompanied by the phase slippage process. At low enough temperatures such fluctuations are solely due to quantum effects. Accordingly, this quantum phase slip (QPS) process can be interpreted as tunneling of the superconducting phase $\varphi$ between the states differing by $2\pi$. The corresponding tunneling amplitude reads \cite{GZ01} 
\begin{equation}
\gamma_{QPS} \sim (g_\xi\Delta/\xi)\exp (-ag_\xi), \quad a \sim 1.
\label{gamma}
\end{equation}
Thus, by varying the parameter $g_\xi $ one can tune the magnitude of both Gaussian fluctuations of the order parameter and QPS effects in superconducting nanowires. The exponential dependence of QPS effects on  $g_\xi $  (\ref{gamma}) was also verified experimentally \cite{BT,Lau,Zgi08}.

Another important dimensionless parameter $g=R_q/Z_{\rm w}$ that also accounts for fluctuation effects has to do with the fact that a superconducting wire can be viewed as a transmission line with impedance $Z_{\rm w}=\sqrt{\mathcal{L}_{\rm kin}/C_{\rm w}}$, where $\mathcal{L}_{\rm kin}$ and $C_{\rm w}$ are respectively the kinetic wire inductance and the geometric wire capacitance. Obviously, the parameter $g$ is not identical to $g_\xi $ being responsible for a complimentary set of fluctuation phenomena in superconducting nanowires. One of them is predicted theoretically \cite{RSZ} and observed experimentally \cite{Kostya} fluctuation-induced smearing of the electron density of states (DOS). This effect may persist even for very large  values of $g_\xi \gg 1$ and it can be interpreted as a result of interaction between electrons inside the wire and an effective dissipative environment formed by sound-like plasmons \cite{MS} propagating along the wire. In particular, it was demonstrated \cite{RSZ} that at $T \to 0$ the gap singularity in DOS disappears completely at $g\leq 2$. 

The parameter $g$ (or $\lambda \equiv g/8$ that will be used on equal footing with $g$ further below) also controls the strength of (logarithmic in space-time) interaction between different quantum phase slips \cite{ZGOZ}. At $T \to 0$ and $\lambda >2$ the interaction turns out to be strong enough and QPS-anti-QPS pairs are formed in the wire which then demonstrates vanishing {\it linear} resistance \cite{ZGOZ}. In this sense, as long as $\lambda >2$ (or, equivalently, $g >16$) the ground state of the system  can be considered  superconducting. In contrast, for  $\lambda < 2$ inter-QPS interaction is weak, quantum phase slips are unbound, and the wire acquires non-zero resistance which tends to {\it increase} with decreasing $T$. The latter feature allows one to call the wire behavior insulating provided $\lambda <2$. Thus, at $g=16$ and $T \to 0$ one expects a superconductor-to-insulator quantum phase transition to occur in the systems under consideration.

At this stage it is important to emphasize that possible insulating behavior of superconducting nanowires is essentially linked to a certain type of experiment performed with such nanowires and may not always be realized. For example, an ultrathin superconducting nanowire forming a closed ring does not loose the ability to carry supercurrent even for $\lambda <2$. In this case a characteristic length scale \cite{SZ13}
\begin{equation}
L_c \propto \exp \left(\frac{ag_\xi}{2-\lambda}\right), \quad \lambda \leq 2,
\label{Lc}
\end{equation}
emerges beyond which phase coherence (and, hence, supercurrent) gets exponentially suppressed by quantum phase slips \cite{AGZ}.
Obviously, the correlation length  (\ref{Lc}) diverges at $\lambda \to 2$, thus signaling the transition to the ordered phase $\lambda > 2$ with bound QPS-anti-QPS pairs and more robust superconductivity.

On the other hand, one should bear in mind that a closed ring geometry seriously restricts the space for phase fluctuations, thereby enhancing the tendency towards superconductivity, see, e.g., the discussion of this point in Ref. \onlinecite{BFSZ}. For this reason it is highly desirable to analyze ground state properties of superconducting nanowires  where no fluctuation configurations are suppressed by geometry constraints and/or boundary conditions. This task is accomplished in our present work. 

In particular, we will argue that a "disordered" phase $\lambda <2$ (or $g<16$) itself consists of two
different phases: A non-superconducting one with $g<2$ as well as a "mixed" one with $2<g<16$ characterized by two different correlation lengths, $L_c$ (\ref{Lc}) and 
\begin{equation}
L^* \propto g^{\frac{1}{1-2/g}}, \quad g \geq 2.
\label{L*}
\end{equation}
The latter phase demonstrates a non-trivial interplay between supercurrent and quantum fluctuations resulting in superconducting behavior of the wire at shorter length scales combined with its vanishing superconducting response in the long scale limit. 

The structure of our paper is as follows. In Sec. \ref{model} we specify the system under consideration and formulate our theoretical approach based on the effective action technique combined with the self-consistent variational calculation. In 
Sec. \ref{results} we display the results of our calculation for the supercurrent and their analysis in various physical limits. 
Sec. \ref{discussion} contains a discussion of our key observations.

\section{\label{model} Model and formalism}

Let us consider the system depicted in Fig. 1. A long superconducting nanowire with sufficiently small cross section $s$ is attached to two big superconducting reservoirs at its ends. The wire is described by geometric capacitance (per length) $C_{\rm w}$ and kinetic inductance (times length) $\mathcal{L}_{\rm kin}=1/(\pi\sigma_N\Delta s)$, where $\sigma_N$ is the normal state Drude conductance of the wire and $\Delta$ is the (mean field) superconducting order parameter. In order to probe fluctuation effects inside the wire it is connected to a bulk superconductor forming an open ring by two identical small area tunnel junctions with Josephson energy $E_J$ located at a distance $L$ from each other at the points $x=0$ and $x=L$. External magnetic flux $\Phi$  piercing the ring controls the phase difference $\phi=2\pi \Phi/\Phi_0$ between the bulk sides of the point contacts. Fluctuations of the superconducting phase $\varphi (x,\tau)$ remain unrestricted in any point $x$ of the wire. The task at hand is to analyze the effect of these fluctuations on the supercurrent $I(\phi)$ flowing through the wire segment of length $L$ between two Josephson contacts. As $I(\phi)$ is a $2\pi$-periodic function of $\phi$ in what follows it suffices to restrict the phase interval to $|\phi|\leq \pi$.

\begin{figure}[h]
\includegraphics[width=0.99\linewidth]{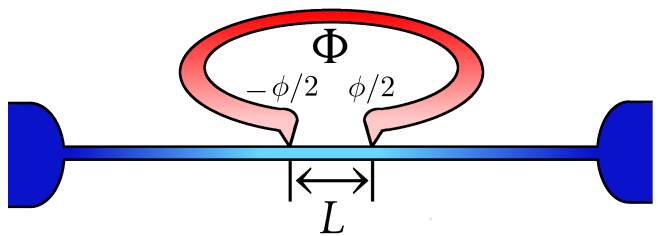}
\caption{The system under consideration.}
\label{FIG1}
\end{figure}

Low energy processes in the above system can be described by the effective action
\begin{equation}
S[\varphi]=S_{\rm w}[\varphi]+S_{J}[\varphi (0), \varphi (L)],
\label{system_action}
\end{equation}
where 
\begin{equation}
S_{\rm w}[\varphi]=\frac{C_{\rm w}}{8e^2}\int\limits_0^{1/T} d\tau \int dx \left[
\left(\frac{\partial\varphi}{\partial \tau}\right)^2+v^2\left(\frac{\partial\varphi}{\partial x}\right)^2
\right]
\label{wire_action}
\end{equation}
is the low energy effective action for a superconducting wire \cite{AGZ,GZ01,ZGOZ,OGZB} and 
\begin{equation}
S_{J}[\varphi_1, \varphi_2]=-E_J\int\limits_0^{1/T} d\tau\bigl[
\cos(\varphi_1+\phi/2)+\cos(\varphi_2-\phi/2)
\bigr]
\label{int}
\end{equation}
accounts for the Josephson energy of the contacts. Here and below we set $\varphi_1=\varphi(0,\tau)$, $\varphi_2=\varphi(L,\tau)$, $e$ is the electron charge and  $v=1/\sqrt{\mathcal{L}_{\rm kin}C_{\rm w}}$ is the velocity of the plasmon mode \cite{MS}.  For simplicity, in Eq. (\ref{int})) we do not include the charging energy of the point contacts which can be absorbed into the first term of the wire action (\ref{wire_action}).

\subsection{Reduced effective action}

As the wire action (\ref{wire_action}) is Gaussian it is possible to exactly integrate out the phase variable $\varphi (x)$ at all coordinate values $x$ along the wire except for the points $x=0,L$. After that we arrive at the reduced effective action $S_R$ for our structure depending only on the two phase variables  $\varphi_1$ and $\varphi_2$. The whole procedure is straightforward, and for the grand partition function $\mathcal{Z}$ we obtain
\begin{align}
\mathcal{Z}&=\int D\varphi(x)\,{\rm e}^{\ts -S_{\rm w}[\varphi(x)]-S_{J}[\varphi(0),\varphi(L)]}\nn\\
&=\int D\varphi_1 D\varphi_2\, {\rm e}^{\ts -S_R[\varphi_1,\varphi_2]-S_{J}[\varphi_1,\varphi_2]},
\label{B}
\end{align}
where
\begin{equation}
S_R[\varphi_1,\varphi_2]=
\frac{1}{2}{\rm Sp}
\begin{pmatrix}
\varphi_1 \\ \varphi_2
\end{pmatrix}^T
\begin{pmatrix}
G_0(0) & G_0(L) \\
G_0(L) & G_0(0)
\end{pmatrix}^{-1}
\begin{pmatrix}
\varphi_1 \\ \varphi_2
\end{pmatrix}
.\label{eq.3}
\end{equation}
Here the trace also includes integration over the imaginary time. The Green function $G_0=\langle \varphi(x,\tau) \varphi(0,0)\rangle_{S_0}$ has the form
\begin{equation}
G_0(\omega_n,x)=\frac{4e^2}{C_{\rm w}}\int\frac{dq}{2\pi}\,\frac{{\rm e}^{\ts iqx}}{\omega_n^2+v^2q^2}
=\frac{4\pi}{g|\omega_n|}{\rm e}^{\ts-\left|
\frac{\omega_n x}{v}
\right|}.
\end{equation}
with $g=2\pi C_{\rm w}v/e^2$. In order to diagonalize  the quadratic part of the action it is convenient to express the reduced effective action in terms of the variables $\varphi_\pm=(\varphi_1\pm \varphi_2)/2$. Then we get
\begin{align}
&S_R+S_{J}
=\frac{1}{2}\sum\limits_{a=\pm}{\rm Sp}\Bigl(
\varphi_aG_{0,a}^{-1}\varphi_a
\Bigr)\nn\\
&-2E_J\int\limits_0^{1/T} d\tau\,\cos\varphi_+ \cos(\varphi_--\phi/2)\label{S_eff}
\end{align}
with the propagators
\begin{equation}
G_{0,\pm}(\omega_n)=\frac{2\pi}{g|\omega_n|}\left(1\pm {\rm e}^{\ts-\left|
\frac{\omega_n L}{v}
\right|}
\right).
\end{equation}

It is worth pointing out that the phase variables $\varphi_+$ and $\varphi_-$ account for different physics in our problem. 
The variable $\varphi_-=(\varphi(L)-\varphi(0))/2$ determines the supercurrent flowing inside the wire segment of length $L$
in-between two contacts. Hence, configurations with non-zero $\varphi_-$ have non-zero energies due the kinetic inductance of the wire and the mode corresponding to $\varphi_-$ has a mass equal to $gv/2\pi L$. In contrast, the variable $\varphi_+$ describes simultaneous shifts of both phases $\varphi(0)$ and $\varphi(L)$ by the same value without producing any phase gradient along the wire, thereby implying that in the absence of interactions the mode corresponding to  $\varphi_+$ is massless. At the same time, below we will observe that fluctuations of $\varphi_+$ yield renormalization of the Josephson coupling energies $E_J$ of the contacts and, hence, should also be taken into account. 

\subsection{Variational analysis}

In order to proceed we will make use of the variational perturbation theory \cite{klnrt}. The key idea of this method is 
to improve the standard perturbation expansion by adding an extra term $\delta S$ depending on the variational parameters to the quadratic part of the action $S_R$. For this purpose the partition function (\ref{B}) can be identically rewritten as
\begin{align}
\mathcal{Z}&=\int D\varphi_1 D\varphi_2\,{\rm e}^{\ts -S_{\rm tr}}\times{\rm e}^{\ts -(S_{J}-\delta S)}
\label{eq.7}
\end{align}
with the trial action $S_{\rm tr}=S_R+\delta S$. The last exponent can then be conveniently expanded in powers of
$S_{J} -\delta S$. If expanded to {\it all} orders, the partition function (\ref{eq.7}) should not depend on the choice of $\delta S$ and the variational parameters.  However, such a dependence emerges provided only a finite number of terms of this expansion is kept. In this case the most accurate approximation is achieved by minimizing the result of the perturbative expansion with respect to the variational parameters.

To this end we choose the trial action in the form
\begin{align}
S_{\rm tr}&=\frac{1}{2}{\rm Sp}\, \varphi_+(G_{0+}^{-1}+m_+)\varphi_+\nn\\
&+\frac{1}{2}{\rm Sp}\, (\varphi_--\psi)(G_{0-}^{-1}+m_-)(\varphi_--\psi), \label{ansatz}
\end{align}
that corresponds to effectively performing a self-consistent harmonic approximation (SCHA). 

Here $m_\pm$ represent the interaction-generated effective masses for the modes corresponding to the phase variables $\varphi_\pm$. The parameter $\psi$ accounts for the average value of the phase difference $(\varphi(L)-\varphi(0))/2$. Note that a somewhat similar variational calculation with a massive term was elaborated in Ref. \cite{FisherZwerger} in the context of Brownian motion of a quantum particle in a periodic potential with linear Ohmic dissipation. The results obtained within the framework of this procedure were found to be in agreement with those derived by means of more rigorous methods \cite{SchZ}. 

Expanding the last exponent in Eq. (\ref{eq.7}) in powers of 
$S_{J} -\delta S$ and evaluating the integrals, for  the free energy $\mathcal{F}=-T\ln \mathcal{Z}$ we obtain
\begin{equation}
\mathcal{F}=\mathcal{F}_0+\mathcal{F}_1+ {\rm higher \;order \;terms},
\label{14}
\end{equation}
where
\begin{align}
\mathcal{F}_0&=\frac{T}{2}\left({\rm Sp}\,\ln\, G^{-1}_+ + {\rm Sp}\,\ln\, G^{-1}_-\right),\\
\mathcal{F}_1&=\bigl\langle S_{\rm int}-\delta S\bigr\rangle_{\rm tr}
\label{16}
\end{align}
Disregarding all higher order terms in the expansion (\ref{14}) and evaluating the average in Eq. (\ref{16})
with respect to $S_{\rm tr}$  (\ref{ansatz}), we get
\begin{align}
\mathcal{F}_1&=-\frac{m_+}{2} G_+(0)
-\frac{m_-}{2}G_-(0)\nn\\
&+\frac{1}{2}\psi\, G_{0-}^{-1}(\omega_n=0)\,\psi\nn\\
&-2E_J\cos(\psi-\phi/2) {\rm e}^{-\bigl(G_+(0)+G_-(0)\bigr)/2}
\end{align}
where $G_\pm^{-1}=G_{0,\pm}^{-1}+m_\pm$. Here and below we denote $G_\pm (0)=T\sum\limits_{\omega_n} G_\pm(\omega_n)$.

Taking variational derivatives of $\mathcal{F}$ with respect to $m_\pm$ and $\psi$ one readily finds
\begin{align}
\label{18}
&\frac{\delta \mathcal{F}_0}{\delta m_\pm}=G_\pm(0)/2,\\
&\frac{\delta \mathcal{F}_0}{\delta \psi}=0,\\
&\frac{\delta \mathcal{F}_1}{\delta m_\pm}=- G_\pm(0)/2
-\frac{1}{2}\frac{\delta G_\pm(0)}{\delta m_\pm}\nn\\
&\times\left(
m_\pm-2E_J\cos(\psi-\phi/2){\rm e}^{-\bigl(G_+(0)+G_-(0)\bigr)/2}
\right),\\
&\frac{\delta \mathcal{F}_1}{\delta \psi}=2E_J\sin(\psi-\phi/2){\rm e}^{-\bigl(G_+(0)+G_-(0)\bigr)/2}\nn\\
&+G_{0-}^{-1}(\omega_n=0)\psi.
\label{21}
\end{align}
Imposing the extremum conditions $\delta\mathcal{F}/\delta m_\pm=\delta\mathcal{F}/\delta \psi=0$ and making use of Eqs.
(\ref{18})-(\ref{21}) we arrive at the following set of SCHA equations:
\begin{equation}
m_+=m_-\equiv m
\label{22}
\end{equation}
and
\begin{align}
&2E_J\cos(\psi-\phi/2){\rm e}^{-\bigl(G_+(0)+G_-(0)\bigr)/2}-m=0,\label{eq_RG}\\
&2E_J\sin(\psi-\phi/2){\rm e}^{-\bigl(G_+(0)+G_-(0)\bigr)/2}+\frac{gv}{2\pi L}\psi=0.\label{eq_mot}
\end{align}
Note that the masses $m_\pm$ in Eq. (\ref{22}) turn out equal because of the symmetry of our problem (i.e. identical Josephson junctions). Equation (\ref{eq_RG}) establishes the relation between the effective mass $m$ and the fluctuation-induced renormalization of the Josephson coupling energy $E_J$. Equation (\ref{eq_mot}) is just the equation of motion for $\psi$. It coincides with the equation of motion for $\varphi_-$ with $E_J$ renormalized by fluctuations.

\subsection{Propagators}

As we already indicated, the wire effective action in the form (\ref{wire_action}) applied only in the low energy limit, i.e. for $\omega, vq \ll \Delta$. Hence, a proper ultraviolet cutoff should be introduced in our calculations which respects both causality and the fluctuation-dissipation relation. This goal is achieved by modifying the spectral density 
$$J_\pm(\omega)=-\frac{1}{\pi}{\rm Im}\, G^R_\pm(\omega)$$ 
making it decay at $\omega>\Delta$. The retarded Green function $G_\pm^R(\omega)$ can be obtained from its Matsubara counterpart by means of the standard analytic continuation procedure
$$G_\pm^R(\omega)=-G_\pm(i\omega_n)|_{i\omega_n\rightarrow \omega+i0}, \;\;\;\omega_n>0.$$ 
Then the Matsubara frequency summation in $G_\pm(0)$ can be performed by means of the contour integration in the complex plane.

Making use of our regularization procedure one finds
\begin{align}
G_\pm(0)&=T\sum\limits_{\omega_n}G_\pm(\omega_n)=\frac{2\pi}{g}T\sum\limits_{\omega_n}\left(
\frac{|\omega_n|}{1\pm{\rm e}^{-|\omega_n L/v|}}+\mu
\right)^{-1}\nn\\
&=\frac{1}{4\pi i}\int\limits_\mathcal{C}dz\, G_\pm(-iz)\coth\frac{z}{2T}\nn\\
&=\frac{i}{4\pi}\int\limits_{-\infty}^\infty d\omega \Bigl(G^R_\pm(\omega)-G^A_\pm(\omega)\Bigr)\coth\frac{\omega}{2T}\nn\\
&=\int\limits_0^\Delta d\omega\, J_\pm(\omega)\coth\frac{\omega}{2T},
\end{align}
where the spectral density functions $J_\pm(\omega)$ read
\begin{align}
&J_\pm(\omega)=-\frac{1}{\pi}{\rm Im}\left[\frac{2\pi}{g}
\left(
\frac{i\omega}{ \Bigl(1\pm {\rm e}^{i\omega L/v}\Bigr)}-\mu
\right)^{-1}
\right],
\end{align}
where  $\mu=2\pi m/g$. In the limit $\mu L/v\ll 1$ these expressions reduce to 
\begin{equation}
J_+(\omega) =\frac{4}{g}\frac{\omega}{\omega^2+4\mu^2}, \quad J_-(\omega) =0.
\end{equation}

\section{\label{results} Supercurrent}

\subsection{Quantum phase transition}
Now we are ready to evaluate the supercurrent $I$ flowing in the wire segment of length $L$ in-between two Josephson junctions. It reads
\begin{align}
I&=-2eT\frac{1}{\mathcal{Z}}\frac{d\mathcal{Z}}{d\phi}=2e \frac{d \mathcal{F}}{d\phi}\nn\\
&=-2eE_J\sin(\psi-\phi/2){\rm e}^{-\bigl(G_+(0)+G_-(0)\bigr)/2}\nn\\
&= \frac{gev}{2\pi L}\psi.\label{current}
\end{align}
Thus, within the accuracy of our calculation the effect of phase fluctuations boils down to
effective renormalization of the critical current $2eE_J$ by the factor ${\rm e}^{-\bigl(G_+(0)+G_-(0)\bigr)/2}$. 

We will further restrict our analysis to the zero temperature limit $T\to 0$. In this case the solution of Eq. (\ref{eq_RG}) 
takes the form
\begin{equation}
\mu=\left\{
\begin{matrix}
\Delta \left(
\frac{4\pi E_J\cos(\psi-\phi/2)}{g\Delta}
\right)^{\ts\frac{g}{g-2}}, & g>2, \\
0, & g< 2,
\end{matrix}
\right. 
\end{equation}
while the renormalized equation of motion (\ref{eq_mot}) can be rewritten as 
\begin{equation}
\frac{\mu L}{v} \tan(\psi-\phi/2)+\psi=0 \label{rewr_eq_mot}.
\end{equation}
It follows immediately that for $g< 2$ we have $\psi=0$ and, hence, the supercurrent $I$ inside the wire 
is fully suppressed by quantum fluctuations of the phase. In contrast, at bigger values of $g>2$ a non-vanishing supercurrent $I$ can flow across the wire segment between two contacts. 

Thus, we arrive at an important conclusion: A quantum phase transition (QPT) occurs at $g=2$ separating two different phases with non-superconducting ($g< 2$) and superconducting-like ($g>2$) behavior. This dissipative QPT belongs to the same universality class as the so-called Schmid phase transition in resistively shunted Josephson junctions \cite{SchZ}. It is curious that this QPT occurs at exactly the same value of the parameter $g$ where the superconducting gap singularity in the local electron density of states gets suppressed due to interaction between electrons and a dissipative bath formed by Mooij-Sch\"on plasmons \cite{RSZ}. We also note that a somewhat similar QPT was also discussed for a single Josephson junction embedded in a thin superconducting ring \cite{HG}.

\subsection{Current-phase relation in the presence of quantum fluctuations}

In what follows we will merely address the properties of a superconducting-like phase $g>2$ and evaluate the supercurrent
$I$ affected by quantum fluctuations of the superconducting phase $\varphi$ in the wire. For this purpose let us combine
the solution of the equation
\begin{align}
\frac{\Delta L}{v} \left(
\frac{4\pi E_J}{g\Delta}
\right)^{\ts\frac{g}{g-2}}\sin(\psi-\phi/2)[\cos(\psi-\phi/2)]^{\frac{2}{g-2}}+\psi =0
\end{align}
with Eq. (\ref{current}).  We immediately observe that there exists a new length scale $L^*$ in our problem associated with the effective mass. Introducing the dimensionless normal state conductance of tunnel junctions $g_N$ and making use of the Ambegaokar-Baratoff formula for the Josephson coupling energy $E_J=g_N\Delta/8$ we express $L^*$ in the form 
\begin{equation}
L^*= \frac{v}{\Delta} \left(
\frac{2g}{\pi g_N}
\right)^{g/(g-2)}
\label{L*}
\end{equation}
We observe that the length scale (\ref{L*}) depends on the relation between two dimensionless parameters: $g$ and $g_N$. 
Here we are merely interested in the case of small-size tunnel junctions with few conducting channels which just serve as probes aiming to disturb the superconducting wire as little as possible. Accordingly, the parameter $g_N$ needs to be small and one typically has $g_N \ll g$. In this case $L^*$ diverges at $g \to 2$ remaining much longer than the characteristic length scale $v/\Delta$ at any value of $g > 2$. 

The length scale  (\ref{L*}) separates two different fluctuation regimes. For $L\gg L^*$ the wire kinetic inductance contribution remains too small as compared to that of the contacts. In this limit the phase difference across the wire segment in-between the contacts is fixed to be $\varphi(L)-\varphi(0)=\phi$ and it does not fluctuate. Then we immediately arrive at the standard mean field current-phase relation 
\begin{equation}
I(\phi)=\frac{gev}{4\pi L}\phi.
\label{mf}
\end{equation}
Our interest, however, is merely focused on the opposite limit  $L\ll L^*$, for which the renormalization of $E_J$ becomes important and phase fluctuations tend to suppress the supercurrent flowing across the wire. In this limit we arrive at the $L$-independent result
\begin{equation}
I(\phi)=\frac{gev}{2\pi L^*}\sin\frac{\phi}{2}\left[\cos\frac{\phi}{2}\right]^{\frac{2}{g-2}}.
\label{I(T=0)}
\end{equation}

Comparing the expressions (\ref{mf}) and (\ref{I(T=0)}) we observe that quantum fluctuations of the phase can strongly
affect both the magnitude and the phase dependence of the supercurrent. The dependence $I(\phi)$ (\ref{I(T=0)}) in the presence of fluctuations becomes smoother than in Eq. (\ref{mf}) and the absolute value of the supercurrent is reduced 
by the factor $\sim L/L^*$. Extra -- phase dependent -- suppression of $I$ originates from the term in the square brackets 
in Eq. (\ref{I(T=0)}):  With increasing $\phi$ the supercurrent gets suppressed stronger and stronger. The latter effect becomes particularly significant for $g$ sufficiently close to 2. For $\phi\rightarrow \pi$ and any $g>2$ the supercurrent tends to zero as $I(\phi) \propto (\pi -\phi)^\frac{g}{g-2}$. 

\begin{figure}[h]
\includegraphics[width=0.99\linewidth]{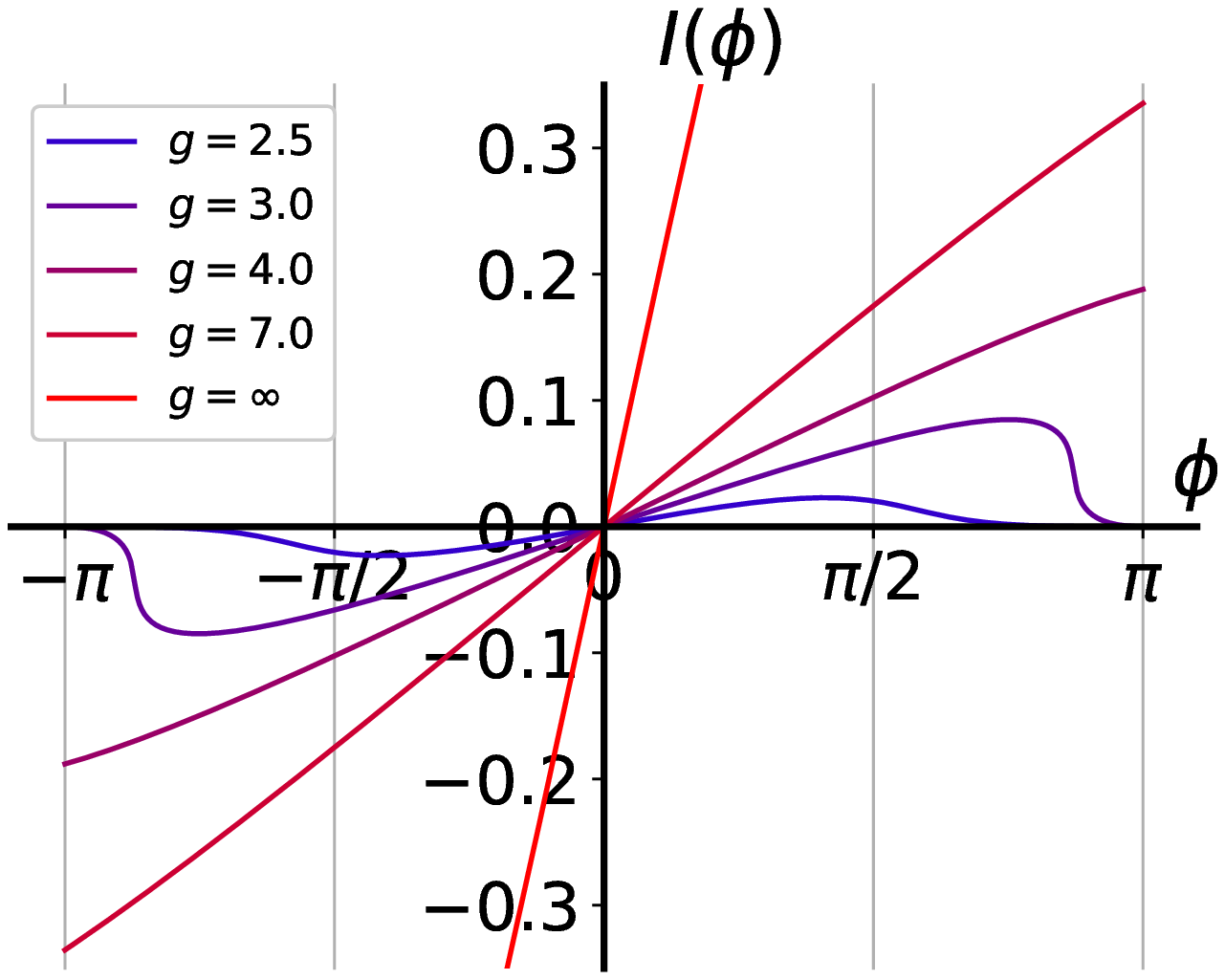}
\includegraphics[width=0.99\linewidth]{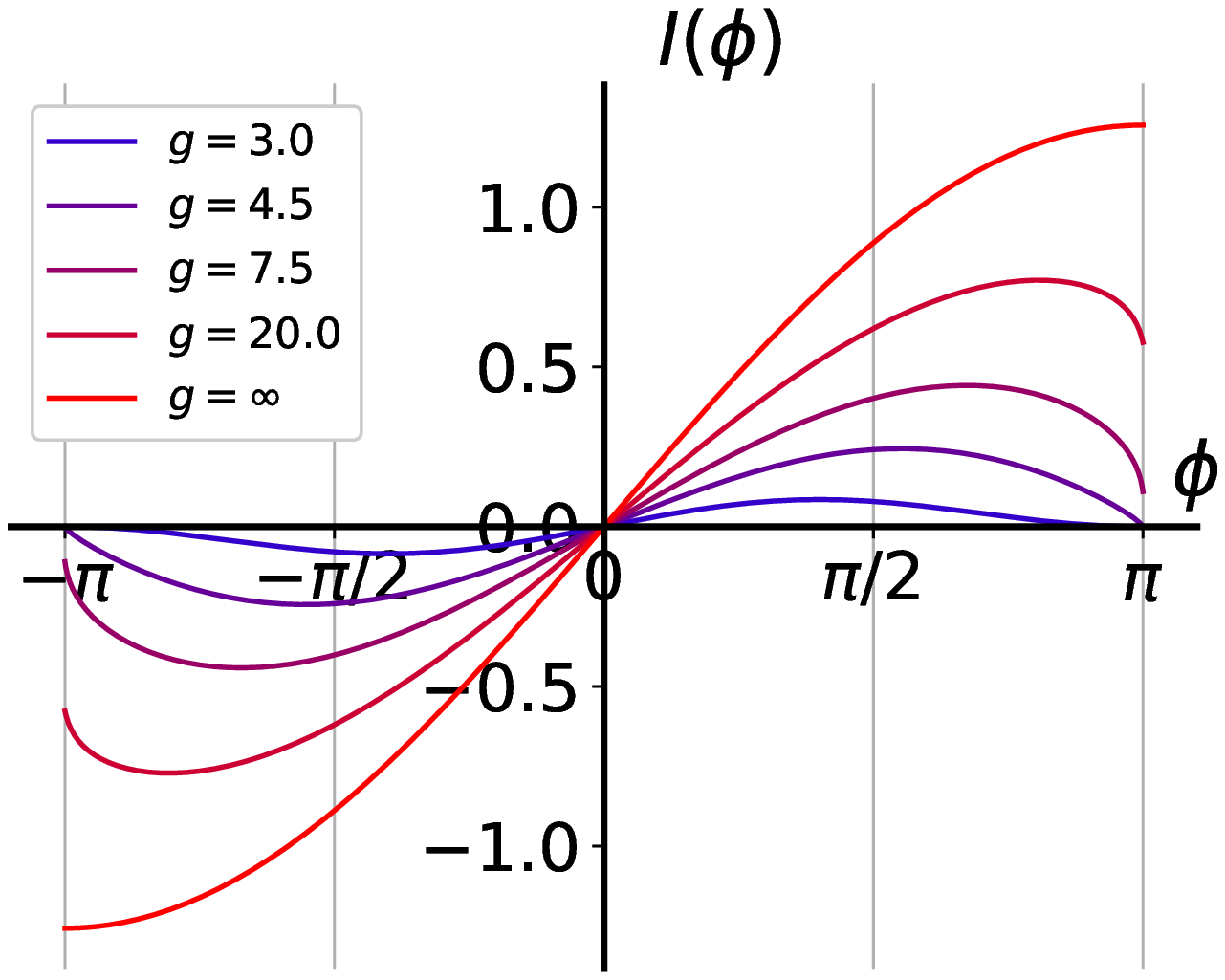}
\caption{The phase dependent supercurrent $I(\phi)$ (expressed in units of $e\Delta/(2\pi)$)  for $E_J/\Delta=0.1$ and different $g$. The upper and lower panels correspond respectively to $\Delta L/v=20$ and to $\Delta L/v=0.5$.}
\label{FIG2}
\end{figure}

Let us also point out that for $L$ not much smaller than $L^*$ the supercurrent  $I( \phi \to \pi)$ behaves somewhat differently: It vanishes only for $2<g<4$, whereas at $g>4$ we have 
\begin{equation}
I( \phi \to \pi)\approx \frac{ge\Delta}{2\pi }\left(\frac{\Delta L}{v}\right)^\frac{2}{g-4}\left(\frac{\pi g_N}{2g}\right)^\frac{g}{g-4},
\end{equation}
i.e. for such values of $g$ the current-phase relation remains discontinuous at $\phi=\pi$. The dependencies $I(\phi)$ evaluated for different values of $g$ and $L$ are also displayed in Fig. \ref{FIG2}. 

Finally, we note that the form of the current-phase relation (\ref{I(T=0)}) obtained here resembles that derived for resistively shunted Josephson junctions in the presence of quantum fluctuations of the phase \cite{PZ}.

\subsection{Effect of QPS}

It is important to point out that the above analysis only accounts for the effect of Gaussian fluctuations of the superconducting phase and does not yet include quantum phase slips. In order to describe QPS effects inside the wire it is convenient to turn to the so-called dual representation for the wire effective action \cite{SZ13}
\begin{align}
\tilde S_{\rm w}&=\frac{1}{\pi g v}\int\limits_0^{1/T} d\tau\int dx\, \left[
\left(\frac{\partial\chi}{\partial \tau}\right)^2+v^2\left(\frac{\partial\chi}{\partial x}\right)^2\right]\nn\\
&-\gamma_{QPS}\int\limits_0^{1/T} d\tau\int dx\, \cos\chi ,
\label{SG}
\end{align}
where $\gamma_{QPS}$ is the QPS amplitude defined in Eq. (\ref{gamma}). This effective action is expressed in terms of the quantum field $\chi (x,\tau )$ which determines electric charge that has passed through the point $x$ of the wire up to moment $\tau$. The quantum operator $\hat \chi (x)$ corresponding to this dual variable obeys the commutation relations 
 \begin{equation}
[\hat\Phi(x),\hat\chi(x')]=-i\Phi_0\delta(x-x'), 
\end{equation}
where $\hat\Phi(x)=\nabla\hat\varphi(x)/2e$ is the flux operator.

The action (\ref{SG}) defines an effective sine-Gordon model which has a QPT at $T \to 0$ and $\lambda=2$ (or $g=16$)
separating two different phases \cite{ZGOZ}. Provided $g>16$ "positive" and "negative" quantum phase slips are bound in close "neutral"  pairs which do not disrupt phase coherence at any relevant scales exceeding the superconducting coherence length $\xi$. Hence, for such values of $g$ QPS effects are irrelevant for any of the above results for the supercurrent $I(\phi)$ which remain applicable without any modifications. 

On the other hand,  for $g<16$ quantum phase slips are no longer bound in pairs. In this phase relevant excitations of our theory are kinks and anti-kinks with characteristic masses $M_s\propto \Delta(v\gamma_{QPS}/\Delta^2)^{1/(2-\lambda)}$ as well as their bound states \cite{LZ,GNT}. The appearance of a gap in the spectrum for $g<16$ (or $\lambda <2$) gives rise to the correlation length \cite{SZ13}
\begin{equation}
L_c\sim \frac{v}{M_s}\sim \xi \exp\left(\frac{ag_\xi}{2-\lambda}\right)\left(\frac{\xi \Delta}{v}\right)^{\frac{1}{2-\lambda}}.\label{LQPS}
\end{equation}
In the context of a setup considered here this correlation length is of little relevance for $g<2$ since in this case the supercurrent
$I$ is totally suppressed already by smooth phase fluctuations. 

On the other hand, for $2<g<16$ the length scale (\ref{LQPS}) becomes important. Actually, for such values of $g$ we have two correlation lengths, $L^*$ and $L_c$ defined respectively in Eqs. (\ref{L*}) and (\ref{LQPS}). The first of these lengths diverges at one of the phase boundaries $g=2$ whereas the second one tends to infinity at another phase boundary $g=16$. Comparing the length  $L$ with each of these two correlation lengths we arrive at the conclusion that the phase with intermediate values
of $g$ ranging from 2 to 16 is described by several different regimes.

Let us first consider the situation with $L^*<L_{c}$, in which case there exist three regimes. At $L<L^*$ the supercurrent is strongly affected only by smooth phase fluctuations and not by QPS. In this regime is determined by Eq. (\ref{I(T=0)}). At $L^*<L<L_c$ the supercurrent is practically insensitive to any kind of phase fluctuations and, hence, it is given by a simple mean field formula (\ref{mf}). Finally, for $L > L_c$ the supercurrent gets exponentially suppressed by quantum phase slips and we have \cite{SZ13}
\begin{equation}
I(\phi) \sim    \frac{eg_\xi\Delta \sqrt{L}}{\sqrt{\xi}}\left(\frac{v}{L \Delta}\right)^{\frac{3\lambda}{4}} \exp\left(-\frac{3ag_\xi}{4}-\frac{L}{L_c}\right)\sin\phi .
\label{expsup}
\end{equation}
Obviously, in practical terms the latter regime can be considered non-superconducting provided $L$ strongly exceeds $L_c$.
 
In principle it is also possible to realize the opposite situation with $L^*>L_{c}$, in particular for values of $g$ close to 2. In this case the length $L^*$ becomes of little relevance, and one can distinguish only two regimes: $L<L_c$ and $L>L_c$. The first
one is again superconducting with the supercurrent $I(\phi)$ decreased by smooth phase fluctuations according to Eq. (\ref{I(T=0)}), whereas the second regime corresponds to exponential suppression of the supercurrent by proliferating QPS, cf.
Eq. (\ref{expsup}). No room for the mean field regime (\ref{mf}) exists at $L^*>L_{c}$.

\section{\label{discussion} Discussion}

According to the well known theorem \cite{alro}  the true long range order cannot be established in infinite low dimensional systems as it gets destroyed by fluctuations. This general theorem, however, does not yet allow one to make any conclusion about the presence or absence of superconductivity in any finite-size structure that can be examined in any realistic experiment. Moreover, superconducting properties of low dimensional structures in the presence of quantum fluctuations may significantly depend on particular experimental realization testing such properties. Here we investigate superconducting fluctuations in long quasi-one-dimensional metallic wires by means of a setup displayed in Fig. 1. This setup enables one  to pass an equilibrium supercurrent across a wire segment of an arbitrary length $L$ without restricting phase fluctuations inside the wire by any means.

The physics of our system is determined, on one hand, by an interplay between collective sound-like plasma excitations and the interaction induced by Josephson point contacts and,  on the other hand, by quantum phase slips. The effective bath of collective excitations can be described in terms of two modes, one of which turns out to be massless or, more precisely, Ohmic at low frequencies. We found that the massless mode renormalizes the Josephson coupling energy of the attached contacts, thereby reducing the supercurrent flowing across the wire segment between these contacts. The Ohmic nature of the effective bath naturally yields a Schmid-like dissipative QPT at $T \to 0$ and $g=2$. Another QPT of Berezinskii-Kosterlitz-Thouless type occurs at $T \to 0$ and $g=16$ being controlled by quantum phase slips \cite{ZGOZ}. Observing to the condition $g \propto \sqrt{s}$, just by tuning the wire cross section one can realize both these QPT in the system under consideration.

According to our results, the zero temperature phase diagram of a superconducting nanowire consists of three different phases.
Thicker wires with $g>16$ show a superconducting behavior, albeit with the supercurrent possibly strongly reduced by smooth (Gaussian) quantum fluctuations of the phase, cf. Eq. (\ref{I(T=0)}). In thinnest wires with $g<2$, in contrast, the supercurrent is totally suppressed by quantum fluctuations, i.e. the phase $g<2$ is clearly non-superconducting. It is also remarkable that the superconducting gap singularity in the local electron density of states gets destroyed by interactions between electrons and a dissipative bath of Mooij-Sch\"on plasmons at exactly the same value of the parameter $g=2$ \cite{RSZ}.

Most interesting is the intermediate phase with $2<g<16$ which is characterized by two different correlation lengths  (\ref{L*}) and (\ref{LQPS}) and demonstrates mixed properties depending on the relation between $L$ and these two lengths. Although formally this phase can still be viewed as superconducting for any finite $L$,  a non-vanishing supercurrent can persist only provided $L$ does not exceed $L_c$. For longer wire segments the supercurrent is exponentially suppressed due to QPS (cf. Eq. (\ref{expsup})). Hence, in practical terms the mixed phase with $2<g<16$ is characterized by a superconducting behavior
of the wire at shorter scales not exceeding $L_c$ and a non-superconducting one at longer length scales.

Perhaps we can also add that fluctuations of the superconducting phase -- both Gaussian and non-Gaussian -- yield 
not only a reduction of the absolute value of the supercurrent but may also essentially modify the current-phase relation $I(\phi)$. These modifications become progressively more pronounced for bigger values of the phase $\phi$, see, e.g.,  Eqs. (\ref{I(T=0)}),
(\ref{expsup}) and Fig. 2.

All our predictions can be directly tested in modern experiments with superconducting nanowires.

\vspace{0.5cm}

\centerline{\bf Acknowledgements}

This work was supported in part by RFBR Grant No. 18-02-00586.

\end{document}